\def\be{\begin{equation}}
\def\ee{\end{equation}}
\def\ba{\begin{eqnarray}}
\def\ea{\end{eqnarray}}
\def\nen{\nonumber}
\def\half {{1\over 2}}
\def\Tau {{\cal T}}
\def\figloc#1#2{\epsfysize=3in
    \centerline{\epsfbox{fig#1.ps}}
    \centerline{Figure #1}
    {\raggedright\it   #2 } 
    \bigskip
    }

\documentstyle[preprint,aps, epsf]{revtex}
\tightenlines
\begin{document}
\title{Dumb Holes and the Effects of High 
Frequencies on Black Hole Evaporation}

\author{W. G. Unruh}
\address{ CIAR Cosmology Program\\
Dept. of Physics\\
University of B. C.\\
Vancouver, Canada V6T 1Z1\\
email: unruh@physics.ubc.ca}
 
\maketitle

~

~

\begin{abstract}

The naive calculation of black hole
 evaporation makes the thermal emission depend 
on the arbitrary high frequency behaviour
 of the theory where the theory is certainly
 wrong. Using the sonic analog to black
 holes-- dumb holes-- I show numerically 
that a change in the dispersion relation at high
 frequencies does not seem to alter the evaporation 
process, lending weight to the reality of the
 black hole evaporation process. I also
 suggest a reason for the insensitivity of the 
process to high frequency regime.
\end{abstract}

\section{Dumb Holes}
Black hole evaporation\cite{Hawking} was one
 of the most surprising predictions of the field of
quantum field theory in curved backgrounds.  
Since the phenomenon has not
 been observed experimentally, it is of crucial 
importance that the assumptions
 underlying the prediction be examined 
especially carefully so as to try to understand the 
process as deeply as possible.

One of the most unsettling features of the derivation is the 
dependence of the  derivation on the behaviour
 of the fields at arbitrarily high frequencies. In fact, if we trace,
  in the usual derivation, the origin of the thermal emission
 from a solar mass sized black hole a second after the formation
 of the black hole, that thermal radiation has its origin in frequencies
 in the incoming (vacuum) radiation   of the order of $e^{10^4}$
 (where any units make only trivial changes in the exponent). 
 Since $\hbar \omega$ for such frequencies correspond to 
masses vastly larger that the mass of the universe, it seems
 certain that the quantum gravitational effects of such frequencies
 would completely alter the behaviour of the field at such scales.
 Would such quantum gravitational effects also destroy the thermal
 emission from the black hole?

Since we have no quantum theory of gravity, this question is 
difficult to answer. However, the thermal emission is not only
 characteristic of black holes, it is also characteristic of dumb holes,
 the sonic analog of black holes\cite{Unruh}\cite{Jacobson}. 
A dumb hole forms when the
 velocity of the fluid exceeds the velocity of sound at some 
closed surface. That surface forms a sonic horizon, an exact
 sonic analog of a black hole horizon. As I showed in 1981, 
the propagation of sound waves in such a hypersonic fluid 
flow are exactly
 the same as the propagation of scalar waves in a black hole 
spacetime.

Since the derivation is not commonly known, it worthwhile 
repeating it here.
The isentropic equation for fluid flow in a medium which is has an 
equation of state $p=p(\rho)$ is given by
\ba
\rho \left(\dot {\bf v} +{\bf v}\cdot \nabla {\bf v} \right) = -\nabla 
p(\rho)
\cr
\dot\rho + \nabla(\rho{\bf v})=0
\ea

I will assume that the flow is irrotational, so that we can write 
\be
{\bf v}= \nabla \Phi
\ee
The first equation can then be written as 
\be
\nabla \left( \dot \Phi + {1\over 2} {\bf v\cdot v} + h(\rho)\right) = 
0
\ee
where $h(\rho) = \int {dp\over \rho} $, from which we get Bernoulli's equation
\be
\dot \Phi + {1\over 2}{\bf v\cdot v} +h(\rho) =0
\ee
(The time dependent constant of integration can be 
absorbed into $\Phi$).
We now introduce the small perturbations of $\Phi$ 
and $\rho$ about the background flow, defining 
$\phi=\delta\Phi$ and $\psi={\delta \rho\over \rho}$.
 We get the equations
\ba
\dot\phi + {\bf v}\cdot\nabla \phi + c^2 \psi =0
\cr
\dot\psi +{\bf v}\cdot \nabla \psi +\nabla^2\phi 
+\nabla ln(\rho)\cdot\nabla\phi=0 
\ea

where $c^2 = { \rho} {dh\over d\rho}= {dp\over d\rho}$.
 Dividing the first equation by $c^2$ and adding taking
 the derivative with respect to 
${d\over dt} +{\bf v}\cdot\nabla+\nabla\cdot {\bf v}$ we
 finally obtain
\be
{1\over \rho}\left({d\over dt} +{\bf v}\cdot\nabla 
+ \nabla\cdot {\bf v}\right) {\rho\over c^2}
 \left( {d\over dt} +{\bf v}\cdot\nabla \right) \phi 
- {1\over \rho} \nabla \cdot \rho\nabla \phi=0
\ee
Now, this is exactly the equation of motion 
for a scalar wave in the metric

$$
\sqrt(-g) g^{\mu\nu} =
 \rho\left(\matrix{ c^{-2} & {v^i\over c^2}\cr {v^j\over c^2} & {v^iv^j\over 
c^2} 
- \delta^{ij}\cr}\right)
$$
or in four dimensions
$$
g_{\mu\nu}= \rho\left(\matrix{c^2-{v^2} & v^i\cr v^j & -\delta_{ij}\cr}\right)
$$
If the flow is stationary, we can define 
the Killing vector
\be
\zeta^\mu=\delta_0^\mu
\ee
and find the Killing Horizon   
at the point where
$c^2=v^2$. If ${v^i\over  {(c^2- v^2)}}$ is an integrable vector, we
can also define
$d\tau= (  dt+ {v^i\over {(c^2- v^2)}}dx^i)$ and  get
the metric
\be
ds^2= \rho ((c^2-v^2)d\tau^2 
- ( \delta_{ij} +{v^iv^j\over c^2-v^2})dx^idx^j
\ee

If we furthermore assume that the flow is radial, ( in which case
 ${v^i\over  {(c^2- v^2)}}$ is always integrable) the metric becomes
$$
ds^2= \rho\left((c^2-v^2) d\tau^2 
- {1\over 1-{v^2\over c^2}} dr^2 - r^2 d\Omega^2\right)
$$
where $d\Omega^2= d\theta^2 +sin^2\theta d\phi^2$ 
the usual angular metric.

By exactly the same arguments as for black holes, 
one expects that this hypersonic flow will produce 
thermal radiation with a temperature given by the
 ``surface gravity" at the black hole horizon, which
 in this case, results in a temperature
\be
T= {1\over 2\pi} {dv\over dr}\vert_{v=c}
\ee

However this prediction is again made on the assumption
 that these sonic equations are valid for arbitrarily short 
wavelengths, a clearly ridiculous assumption. At wavelengths
 shorter than of order the intermolecular spacing, 
the equations for sound waves
 deviate significantly from the low frequency
 continuum fluid approximation. Does this change in the fluid
equations at high frequency alter the prediction of thermal emission? 
If it does, one begins to loose faith in the black hole predictions.
If it does not, perhaps it will give us a clue as to the real origin of 
the 
radiation.

While an exact solution of the fluid equations which fully took into account
the atomic nature of the fluid would be most welcome, I am unable to
supply it. However one of the most
 significant of the deviations from continuum fluid flow is the change 
in the
 dispersion relation for sound waves. Typically at
 wavelengths approaching the inter atomic spacing, the group 
velocity of the sound waves drops to values much 
less than the low frequency value. Can this change 
in the dispersion relation change the prediction of a
 thermal spectrum of sound waves emitted from such a dumb hole?

Thus the model I want to test is a model for sound waves
 within a fluid flow with a sonic horizon, or for scalar waves in
a background metric, in which 
the dispersion relation at high frequencies differs from the
 usual low frequency form. In particular the model will 
contain a high frequency cut off, i.e., a frequency above 
which there are no normal modes for the propagation of 
sound waves. As Jacobson has argued, a naive cutoff leads to not only a 
suppression of the thermal emission, but also a suppression of all 
waves and vacuum fluctuations from the hole. In our model however the cutoff
is to  arises naturally in the theory.

Because I am interested in the effects of changes at high frequencies
to the thermal emission, not in the detailed behaviour of a real fluid, 
I will simplify the model so as to make the solution as easy as possible 
while still retaining the features responsible for the thermal emission 
in the 
naive theory. Thus, to make the calculations easy, I will look at a two 
dimensional model
(i.e., one space and one time) of the propagation of waves
 in a flowing medium and one in which the high frequency
 dispersion relation is non-standard.  I will also assume a
 constant velocity of sound (c=1), and
I will assume that 
the background density $\rho$ is a constant. (Note that although a two 
dimensional background flow cannot have a constant density and 
a varying velocity, this assumption does not change the location 
of the horizon, nor does it significantly alter the propagation 
equation of sound waves  except at the very lowest frequencies.) 
 The model equation of motion of the scalar waves I will use in examining 
the effects of alteration of the high frequency behaviour of the theory 
is

\be
\left({\partial_t} + \partial_x v \right) 
\left( \partial_t + v\partial_x\right)
\phi - F^2(\partial_x) \phi =0
\ee
where the function $F(\kappa)$ is such that
 $F$ is an odd analytic function of $k$, 
 $F(\kappa)=\kappa$ for small $\kappa$ and
 $F(\kappa)=Const$ for large $\kappa$. I.e., 
the dispersion function for the fluid ( taking v=0) is
$$\omega^2 = F^2(k)$$
Note that if $F(\kappa)=\kappa$ for all $\kappa$,
 then we have the simple equation
\be
\left[(\partial_t +\partial_x v)(\partial_t+v\partial_x) 
- \partial_x^2 \right] \phi=0
\ee
for which we can make the change of variables
 $\tau= t+\int {vdr\over (1-v^2)}$ to get
\be
\left[(1-v^2)\partial_\tau^2 - {1\over 1-v^2}\partial_x^2 \right]\phi = 
0
\ee
which has as solution 
\be
\phi(\tau,x) = f(\tau-z) +g(\tau+z)
\ee
where $z= \int {dx\over 1-v^2}$
This is just the equation for a scalar field in a two dimensional black 
hole
spacetime, where near the horizon, we can write 
\be
ds^2=2{dv\over dx}\vert_{v=1} (x-x_0) d\tau^2
 - {dx^2\over 2 {dv\over dx}\vert_{v=1} (x-x_0)} 
\ee

Just as for the black hole spacetime this leads to the 
conclusion that the temperature of the emitted radiation is given by 
\be
T={1\over 2 \pi} {dv\over dx}\vert_{v=1}
\ee

Alternatively, one can use the analytic continuation
 arguments to imaginary $\tau $ time and demand 
regularity of the Green's function at the horizon in the regular Euclidean 
metric
\be
ds^2 = (1-v^2) d\Tau^2 + {1\over 1-v^2} dx^2
\ee
with the identification 
\be
{dv\over dx}\Tau +2\pi = {dv\over dx}\Tau \label{identification}
\ee
to arrive at the same conclusion.

However, as soon as we alter the dispersion function $F$, this
 analyticity argument fails. Under the analytic continuation 
of $\tau= i\Tau$, the equation of motion for the field becomes 
 complex, rather than real. One can draw no 
conclusion from
 the behaviour of the Green's function for this complex metric,
 nor does one have any reason to demand that one make the 
identification of the Euclidean $\Tau$ coordinate of eqn. \protect\ref{identification},
 since the equation of motion no longer has a  metric that 
this identification makes regular.

\section{Numerical Solution}
In order to determine the consequence to the thermal emission 
process of the  altered dispersion relation, we must go back to
 basic principles and calculate the Bogoliubov coefficients for 
the conversion of ingoing positive frequency modes to outgoing
 negative frequency modes. I have carried out this calculation 
numerically for a couple of  dispersion functions $F$.

As is usual, the particle production rate is equal to the overlap 
integral between a negative frequency final mode and a positive
 frequency initial mode.
One can either start with a  positive frequency initial mode, propagate 
this 
through the region containing the black or dumb hole, 
and calculate the negative frequency component of the final mode,
 or start with a  negative frequency final mode and calculate the 
positive frequency initial component of this by propagation backwards in 
time. It turns out that
 the latter is the easier to do numerically. Thus I start with some 
outgoing purely negative frequency wave packet. I then propagate
 it backwards in time toward the horizon  until it is converted 
to an initial pulse, and calculate the positive 
frequency component of that initial packet.

It is here necessary to state, in the case of the non-trivial dispersion 
relation, what I mean by positive and negative frequency.  
 This is important because the  dispersion 
relation implies that the notion of positive frequency  does not refer 
to positive or negative sign of the frequencies {\it  per se}. 
The 
dispersion relation for constant velocity of fluid flow  looks like
\be
\omega_\pm = vk \pm |F(k)| \label{disp}
\ee
Even if $v$ is less than unity, there are values of positive $k$ for which 
$\omega_\pm$ both are 
positive, and values of negative $k$ for which both are positive. 
In particular if $k>k_{c}$, 
where $k_c$ is the solution to $F(k_c)=vk_c$, then both possible values,  
$\omega_\pm$, are positive. ($k_c$ represents the wave vector at which the 
phase velocity of the wave is zero.) Thus, in the sense of the sign of
 $\omega$, there are no positive frequencies at all for $k>k_c$ and no 
negative frequencies at all for $k<-k_c$. However, the term `positive
 frequencies' or `negative frequencies' refers not to the sign of $\omega$ (although there is a
correlation for free fields) but to fixed frequency modes which
 have positive or
negative norm.

The field equations for the simplified two dimensional model for sound 
waves (with constant density) can be derived from the action
\be
I= \half \int |\partial_t \phi + v(t,x) \partial_x \phi|^2 - |(F(\partial_x) 
\phi|^2 dx
\ee
which leads to a momentum
\be
\pi(t,x)= \partial_t\phi + v(t,x)\partial_x\phi
\ee
and an inner product
\be
<\phi,\phi'>=-{i\over 2}\int (\phi^*(t,x) \pi'(t,x) - \pi^*(t,x)\phi'(t,x)) 
dx
\ee
If we assume a solution with a definite frequency and wave number
 $\phi_{\omega,k}(t,x)=\tilde \phi(\omega,k) e^{i\omega t +kx}$ 
and  with constant $v$, we find that 
\be
<\phi_{\omega,k}(t,x),\phi_{\omega,k}(t,x)> 
= \pm F(k) \int |\phi_{\omega,k}|^2 
dx
\ee
where $\omega$ is given by equation \protect\ref{disp}. 
The definition of a positive frequency mode is that the mode have a 
definite frequency and that the inner product be 
positive, and thus that $\pm F(k)$ be positive. This selects the positive 
sign for $k>0$ and the negative for $k<0$ and independent of the sign of 
$\omega$.

One can rephrase the above discussion by stating that one is definig positive
frequency by the sign of the frequency in the co-moving frame. Defining 
$x'=x-vt$, the mode $e^{i(\omega_pm t +k x)} $ becomes 
$e^{i((\omega_pm-kv)t+kx)}= e^{i(\pm|F(k)| t +kx')}$.

In the case of the usual black hole situation, or the case in 
which the dispersion relation is linear for all frequencies for 
 the dumb hole,  we must propagate the packet backward in 
time until it enters the region of black (or dumb) hole formation.
 During that back propagation, the packet is constantly blue shifting,
 so that the frequency of the wave by the time it reaches the formation
 time is absurdly high.  This is clearly not calculable numerically, but
 the process has been well understood analytically
 since Hawking's original calculation. In the case
 in which the dispersion relation is such that $\omega$ 
 becomes a constant
 at high frequencies, however, the calculation can be
 done numerically, and the positive and negative frequency 
overlap integral can be computed numerically.

The dispersion function $F$ is taken to be given by
\be
F(k) = k_0\tanh(\left({k\over k_0}\right)^n) ^{1\over n}
\ee
where $k_0$ is the spatial frequency at which the dispersion
function changes from linear to constant.   
This gives the length scale over which the dispersion 
function leads to  a non-locality in the propagation equations of about 
$2\pi/k_0$. 
 The parameter $n$
determines the sharpness of the transition, with large $n$ implying 
a very fast transition between the two regimes . I have studied what happens
 with $n=1$ and $n=4$ in most detail. 
  A calculation with $n=\infty$ was also
 carried out and the results were essentially the same as those for $n=4$. 
I will first describe the $n=1$ case in some detail, and then briefly
present the results for $n=4$.

 \figloc1{The soft  dispersion function ($n=1$) and corresponding group
velocity (dotted line) at zero fluid velocity. The transition wave number 
$k_0$ is 512.}

A non-linear dispersion function means that the group velocity of the wave 
when the fluid velocity,$v$ is $0$, namely
\be
v_g = {d\omega\over d k},
\ee
is not a constant, but depends on frequency. In figure \protect\ref{fig1}
 is plotted 
$\omega $ and $v_g$, the group velocity, as a functions of $k$ for 
the value of $n=1$.   One can now give a hand waving description
 of the effect of this non-trivial 
dispersion function on the behaviour of a wave packet. I will describe 
( and in the next section analyze) the problem by propagating an 
outgoing packet backward in time.  Consider a long wavelength packet which
is an outgoing packet at late times. It travels outward away from the sonic
 horizon and against the fluid flow. As we go backward in time, it approaches 
nearer and
nearer the horizon, until when it is sufficiently close it begins to blue 
shift exponentially.
 Eventually the frequency becomes
 sufficiently high that the group velocity of the 
wave is less that the velocity of the fluid at the horizon. 
The fluid therefore 
drags this outgoing packet in toward the hole ( or, since we are examining 
the 
system backward in  time, the packet at earlier times is further and further 
from the horizon).  The packet seems to bounce off the horizon because of 
the chage in the relation of the group and the fluid velocity of the 
packet.

The naive picture is however complicated by the fact that the
 blue shift of the wave near the horizon is spatially very non-uniform.
 The front edge of the wave, the edge nearest the horizon, is blue
 shifted to a much larger extent ( at any given time) than
 the trailing edge of the wave. In fact the leading edge is blue shifted
to the final frequency long before the tail of the pulse has come near
the sonic horizon. Thus, it seems dangerous to   simply analyze the process 
either in position space, or in momentum space.

\figloc2{ The velocity profile of the fluid used. The system is
assumed to have periodic boundary conditions. Note that $x\approx .7$
corresponds to the horizon. $x\approx .8$ is a Cauchy horizon.}

However, the numerical calculation seems
 to support this naive picture. In the numerical calculations, I have chosen
my space to have periodic boundary conditions with $x=0$ identified with
$x=1$. This was done primarily because of the difficulty of specifying 
any other boundary conditions with the non-trivial wave equation with its
unusual dispersion relations. The velocity field is chosen to be given 
by the 
function
 \be
v(x)= 1- .2 \tanh\left( 24\cos(2\pi(x-.25))+23\right)
\ee
which is plotted in figure   \protect\ref{fig2}. 
This function, which is constant with value of .8 ( in units where the
low wave number velocity of sound in a still fluid is unity) over most 
of the range, was chosen precisely  to have a broad  range of constant
value with a smooth transition
to a region where $v>1$.
Although the region where $v>1$ is small, this turns out to be
irrelevant because the wave never penetrates that region.
The horizon occurs where $v(x)=1$, which is where $\cos(2\pi(x-.25))= -23/24$, 
or 
$x\approx .75\pm .046$. It is the lower value ($x\approx .7$) which represents 
the horizon, while the other represents an irrelevant Cauchy horizon.

Another value of interest is the derivative of velocity with respect 
to position, ${dv\over dx}$ at the horizon, since this gives the temperature.
 This is given by
\be
{dv\over dx}\vert_{v=1} = .2(48\pi\sqrt(1-(23/24)^2)=.4\pi \sqrt(47)\approx 
8.6
\ee
 giving a temperature of 
\be
T={1\over 2\pi}{dv\over dx}= 1.37.
\ee
(Recall that our units are such that $v_g(k=0)=1$, so that the thermal 
wave number is $k_T\approx 1/T$ and the thermal wave length is  thus approximately 
8, which is longer than the whole spatial domain.

  The final (late in time) wave  packets
 (recall that we will be propagating the packets 
backward in time), chosen to be purely 
negative frequency packets,  are
 thus complex, and are traveling away from the horizon. They are
confined as much as is possible to the region
 where the velocity
 is constant. They now travel (backwards
 in time) toward the
 horizon of the dumb hole, where 
they begin to blue shift.
  Eventually their frequency becomes 
sufficiently high that the group 
velocity drops to less than unity and they 
are dragged backward out of the region of
the hole to the region where the fluid 
velocity is constant. In this region we can again 
analyze the wave into its positive and 
negative frequency parts, and determine the
Bogoliubov transformation quantity for this 
particular wave packet.

This analysis must be carried out in the constant 
velocity regions. Whether or not
a wave is "positive frequency" depends on its 
time dependent behaviour. However,
both because of the finite size of the system,
 and because of the finite time for which one
can reliably carry out the integration, the
 temporal behaviour is difficult to use in
determining whether or not the wave is 
positive frequency. However, in regions where the
fluid velocity is constant, one can directly 
compare the relation between the field and its
first time derivative to determine its positive 
and negative frequency components. We have
\ba
(\partial_t-v \partial_x)\phi=\pi
\nen\\
(\partial_t-v\partial_x)\pi= F^2(\partial_x)\phi
\ea
Fourier transforming under the assumption that $v$ is a constant,
we have
\ba
(\omega_\pm -vk) \tilde\phi_\pm(k)=\tilde\pi_\pm(k)
\nen\\
(\omega_\pm -vk)\tilde\pi_\pm(k)= F^2(k) \tilde\phi_\pm(k)
\ea
We can solve for $\omega$ and choose that
\be
\omega_\pm= vk \pm |F(k)|
\ee
with $\omega_+$ the positive frequency branch  for all values of $k$ (independent 
of
the sign of $\omega_+$).
 
Now, given $\tilde \phi(t,k)$ and $\tilde\pi(t,k)$ we can split it into 
its positive and negative frequency parts by
$\tilde\phi(k)=\tilde\phi_+(k)+\tilde\phi_-(k)$ and 
$\tilde\pi(k)=\tilde\pi_+(k)+\tilde\pi_-(k)$, and the $\pm$ subscript
denotes the positive and negative frequency component.
Solving for the positive frequency part, we get
\ba
\tilde\phi_+(k)= \half\left(  (-i\tilde\pi(k)/|F(k)|
  +\phi(k)\right)
\nen\\
\tilde\pi_+(k)= \half\left((\tilde\pi(k)
  -i|F(k)|\phi(k)\right)
\ea

The procedure now is to set up the initial (late time) wave as 
a purely negative frequency wave,
propagate it backward in time until the
 resultant wave is again entirely (or as much as is possible) in the
 constant velocity regime. The resultant 
wave is then analyzed into
 its positive and negative frequency
 parts, and the inner product of each of these
parts separately is calculated.  Note
 that the inner product of the full wave itself is
conserved, and this can be used as a 
test of the integration routine. We finally get
 the actual Bogoliubov coefficients 
for the wave from the future to the past (and thus also from the past to 
the future) and can compare this with the thermal hypothesis.  

One can  also calculate the expected
 Bogoliubov coefficient under the assumption that
the outgoing mode is thermally distributed, 
as would be expected from the naive application
of the Hawking procedure to the wave. 
 The wave packets used are not plane wave states,
and thus one must integrate the thermal
 factor over the spectrum of the outgoing wave to
find the expected amplitude of the thermal Bogoliubov coefficient.
 It this thermal factor equals the Bogoliubov
 coefficient calculated from
the evolution equations, then the 
thermal radiation for dumb holes hypothesis is validated
even with this non standard high frequency dispersion function.

\figloc3{ The Real (solid line) and Imaginary (dotted line) parts of a 
final pulse emerging 
from the
dumb hole. The dispersion function has $n=1$ and $k_0=512$.}

 \figloc4{ An intermediate stage in the evolution backward in time of the 
real part
of the pulse of figure \protect\ref{fig3}. Note that the front
 of the wave has crashed into 
the horizon,
and been blue shifted to a high frequency pulse whose group velocity is 
less than the fluid
velocity. The dotted line is the same calculation but at half the spatial 
resolution as the solid line.}

\figloc5{ The pulse as it must have been initially to produce the outgoing 
pulse of
figure \protect\ref{fig3}.  Note that it is an outgoing
 pulse but is being dragged in toward 
the horizon by the
inflowing fluid. Again the dotted line is at half the resolution as the 
solid line to give an estimate of the numerical errors.}

Figures 3  gives the real (solid line) and
imaginary (dotted line) parts of the wave packet
at the final time.  In figures \protect\ref{fig4} and \protect\ref{fig5}
 I show  the real part of the
 wave at two separate earlier times. In figure \protect\ref{fig4} 
the wave is just interacting
with the horizon. Part of the wave has been blue shifted (going backwards 
in time) and part is still at the final low frequency. Figure 6 gives the 
incoming wave in the constant velocity region as it must have been in order 
to produce the outgoing wave of
 figure \protect\ref{fig3}.  We note in figure \protect\ref{fig4} 
that the front edge of the
wave has already crashed into the ``horizon" ($v=1$), been
 blue-shifted and begun to be dragged back out of the dumb hole 
by the inflowing (which in the
reverse time direction is outflowing) fluid because its group velocity
 has decreased at the higher frequencies.
At the same time the trailing edge of the pulse is still traveling toward 
the hole.

In figures \protect\ref{fig4} and \ref{fig5} I also show the outcome 
of the numerical evolution for two
different grid spacings (the dotted is at half the
resolution of the solid 
line)
 to give a feeling for the accuracy
of the integration. The key difference is in the
 phase velocity of the wave being dragged out
 of the hole. This is expected, because of
the finite differencing of the term $v\partial_x$ 
term in the equations. 
The centered differencing scheme converts
 this term to $iv \sin(2k\Delta_x)/2\Delta_x$ 
instead of the exact $ivk$. This means that at
large $k$, the effective velocity of the fluid is 
decreased by $sin (2k\Delta_x)/2k\Delta_x$.
 I.e., at high frequencies, 
the effective fluid velocity is less leading to 
a slightly smaller dragging velocity
for the high frequency wave, as is seen in the simulation.

\figloc6{ The spectrum of the final pulse of figure \protect\ref{fig3}. 
Note that it is 
purely positive wave number
and negative frequency.}

\figloc7{ The spectrum of the initial pulse of figure \protect\ref{fig5}. 
The solid line is positive frequency components
while the dotted line is negative frequencies. 
Since positive frequencies correspond to positive
wave number and vice versa, this still represents a purely outgoing wave.}

In figure \protect\ref{fig6} I give the  spectrum
 ( in k space) of the final outgoing wave 
packet. 
This wave has been designed to be  purely negative
 frequency, left traveling and concentrated in the region where the 
velocity is constant. It represents a wave
packet traveling  away from  the hole against the inflowing fluid flow.
 In figure \protect\ref{fig7} we have the spectrum corresponding to figure 
\protect\ref{fig6}
, divided into
positive and negative frequency parts, again
in $k$ space. The dotted line is the negative frequency components
while the solid line represents the positive frequency components. 
We note that to very good accuracy,
the interaction with the horizon has left
the wave as a wave purely traveling to the left 
( even though
 the fluid velocity is now sufficiently high to drag 
the left traveling wave to the right). The positive $k$ modes
are negative frequency while the negative $k$ modes are positive frequency.

\figloc8{ A narrower final pulse but otherwise the same as in Figure 
\protect\ref{fig3}.}

\figloc{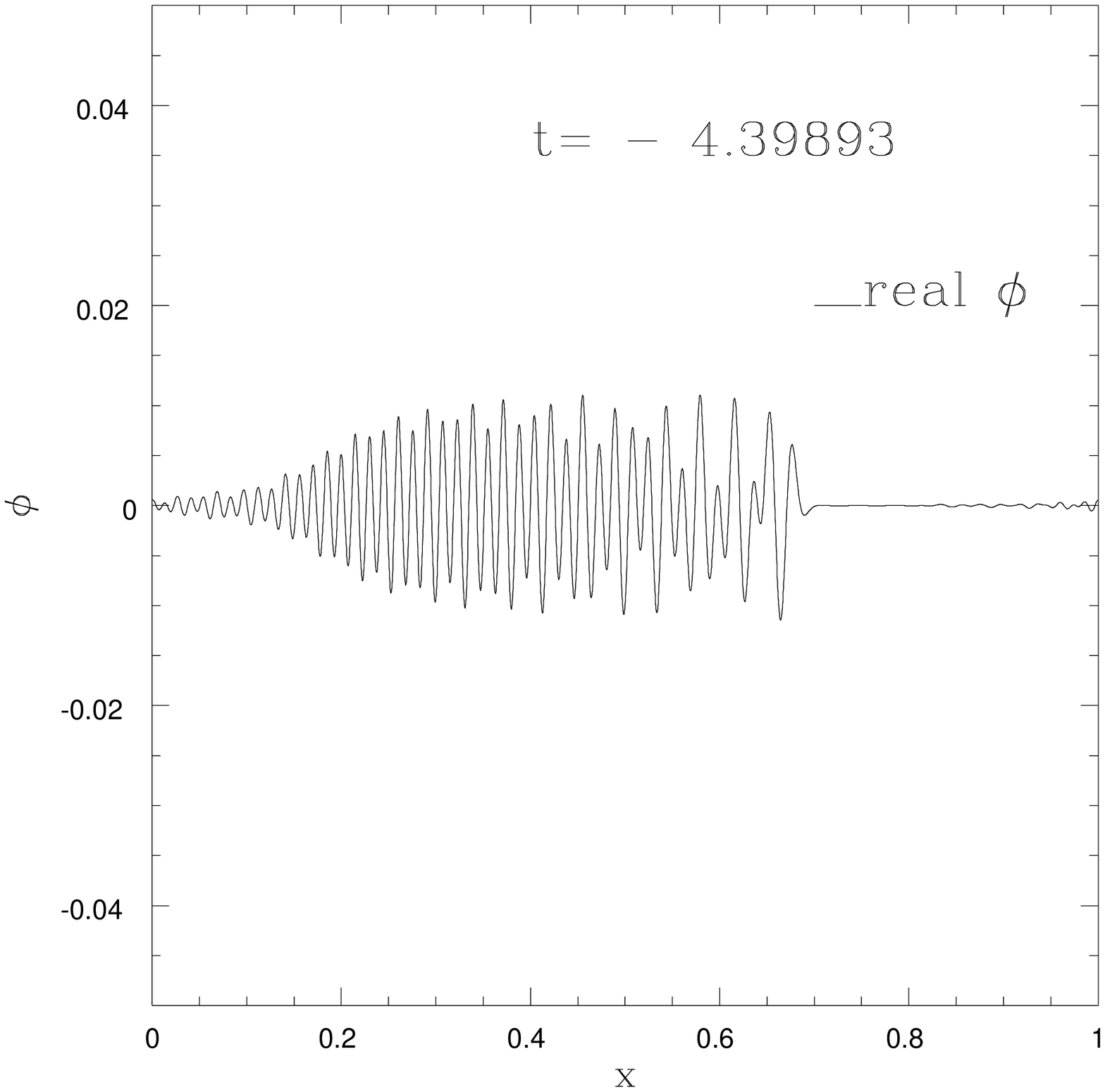}{ The initial pulse which must have produced figure 
\protect\ref{fig8}.}  

\figloc{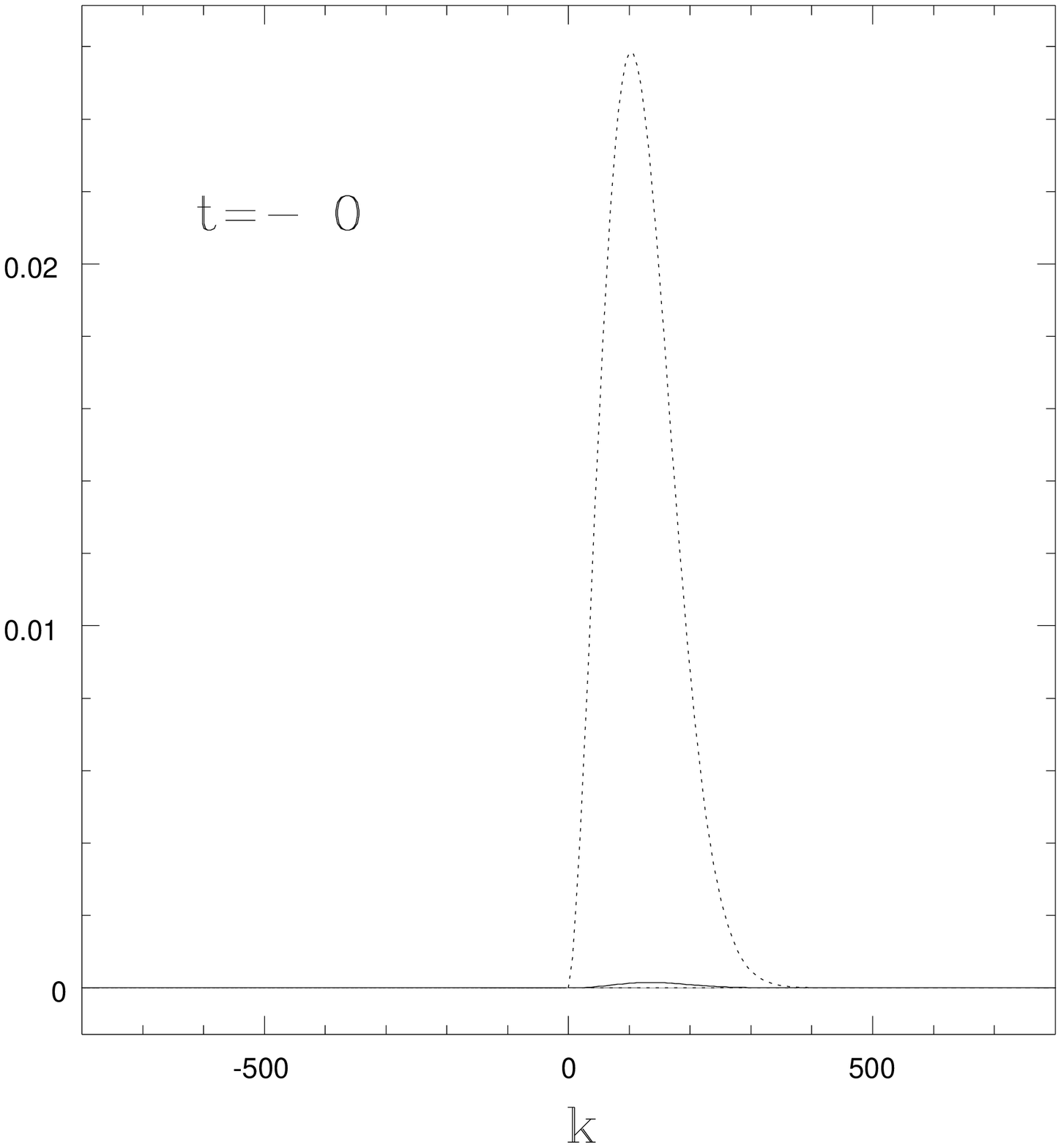}{The spectrum of the final pulse of figure \protect\ref{fig8}.}

\figloc{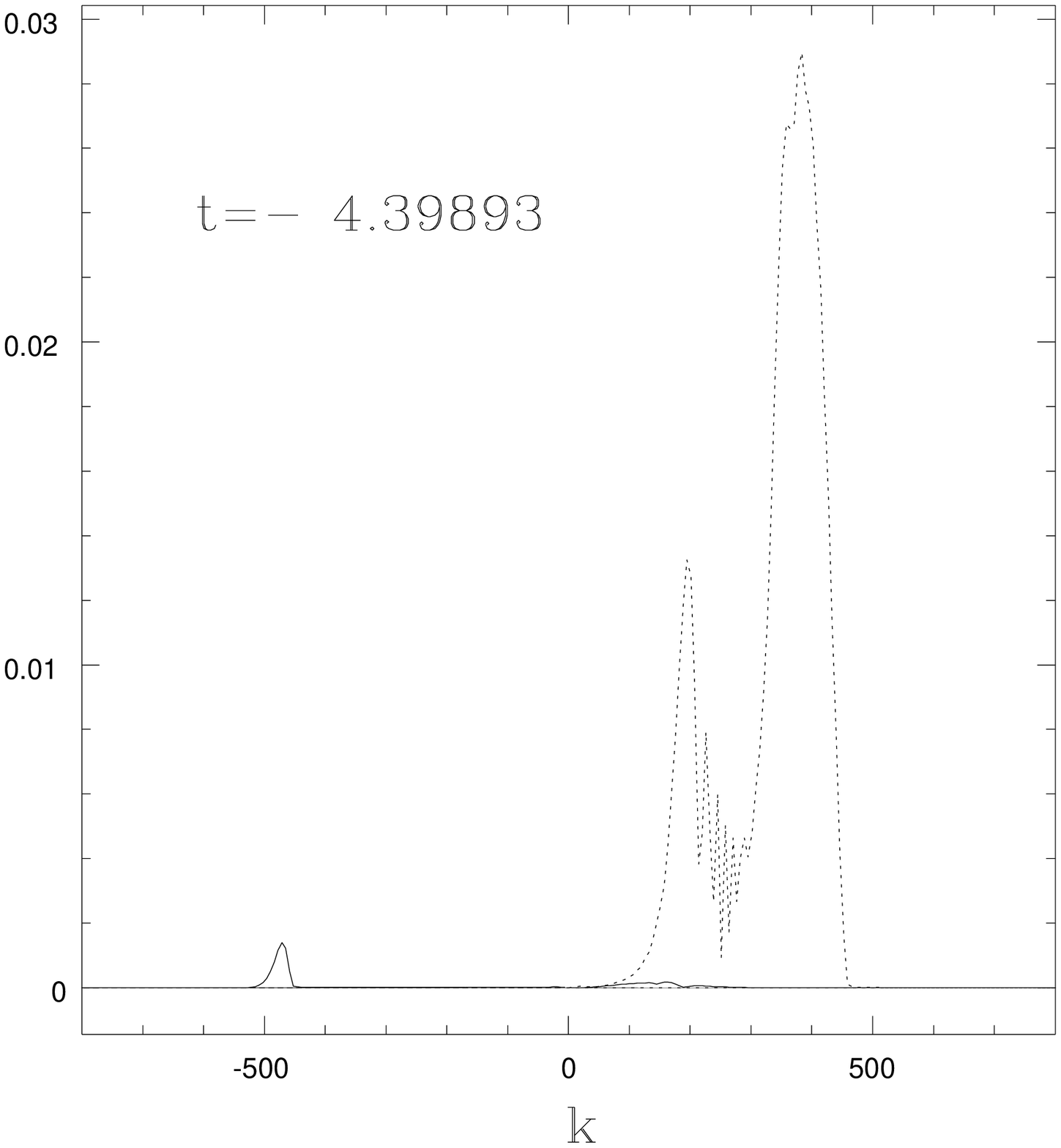}{ The positive and negative frequency spectrum corresponding 
to 
figure
\protect\ref{fig9}.  Note the much smaller positive frequency 
 components than in Figure 
\protect\ref{fig7}. Note that the
power which has not been blue shifted corresponds to components of the 
initial packet whose
group velocity was already of the order of or less than .8, the minimum 
velocity of the
fluid.}

Figures \protect\ref{fig8} and \protect\ref{fig9}
 give the final and initial wave
 packets for another higher frequency pulse
while figures \ref{fig10} and \ref{fig11} give the spectra of these
 pulses again split into positive and negative frequency parts. 
We see in figure \protect\ref{fig11} a much smaller conversion
of positive to negative frequency, as would 
have been expected from the higher frequency of the initial pulse.
All of these cases are for the $n=1$ dispersion function.

Let us now compare the above with the value 
for the total Bogoliubov transformation
expected on the thermal hypothesis.

Given the final mode represented by figure \protect\ref{fig3} 
or \protect\ref{fig8}, the expected number 
of particles resident in this mode in a thermal state of temperature $T$  
is given by
\be
\beta_T={1\over {\cal N}}\int {e^{-\omega T}\over 1- e^{-\omega T}} 
\left(\tilde\pi^*(k)\tilde\phi(k)-\tilde\phi^*(k)\pi(k) \right)dk
\ee
where $\cal N$ is the normalization factor
\be
{\cal N} = \int \left(\tilde\pi^*(k)\tilde\phi(k)
-\tilde\phi^*(k)\pi(k)\right) dk
\ee
On the other hand, the expected number of particles in this mode, $\beta$, 
if we assume that the
input state was in the vacuum state is just given by the positive frequency
portion of the initial wave form ( the Bogoliubov coefficient). 
This is just the norm of the
positive frequency part of the normalized wave.

Comparing the two values, we find for the low frequency pulse in
 the soft ($n=1$ ) dispersion function case  that
\be
\beta_T= .02302; ~~~~~~~~ \beta= .02336
\ee
while for the high frequency pulse
\be
\beta_T=.0005364;~~~~~~~~ \beta=.0005545
\ee
In each case the Bogoliubov estimate of the particle
 production in that particular pulse is essentially the 
same as the thermal estimate. The change in high 
frequency behaviour of the theory makes little or 
no difference to the expected thermal output from
 the dumb hole. ( I suspect the error to be related to the
fact that the final wave pulse is not entirely contained within the
constant velocity region.)  

\figloc{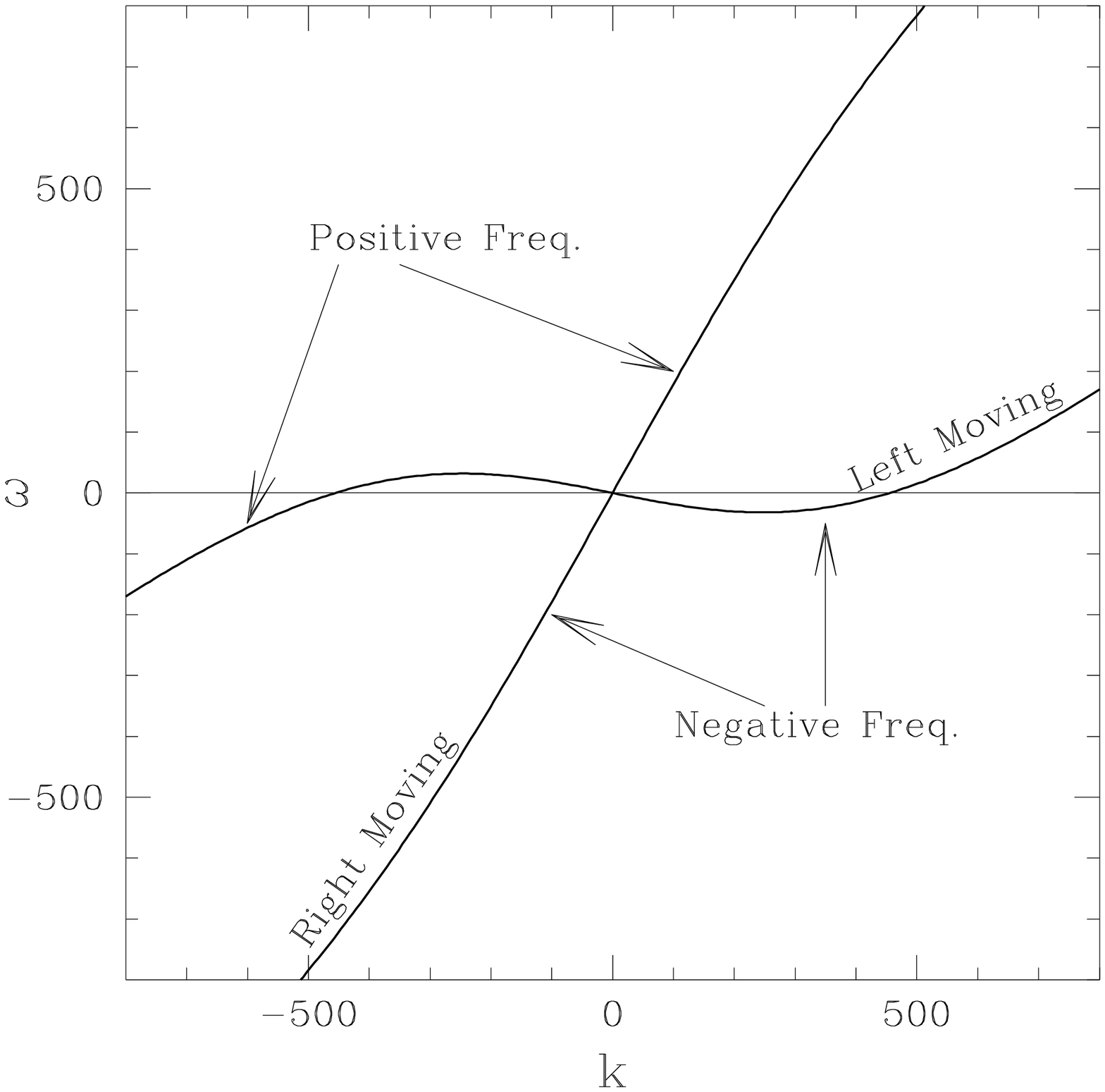}{  The relation between $\omega$, the temporal frequency, and
$k$ the wave number in the constant velocity ($V=.8$) region . The various 
branches  (positive and negative frequency and right and left moving) of 
the dispersion relation are labeled.}

Moreover, in addition to the integrated Bogoliubov transformation, one can 
test whether the full spectrum itself of the outgoing radiation is thermal. 
Since
the 
velocity in the simulation is time independent, the frequency of any wave 
in the
continuum, infinite spatial section case is conserved.  Consider an final 
left moving pulse, contained entirely in the constant velocity region, 
with components
$\tilde \phi(k)$. For any value of k, there are two possible
 values for the frequency $\omega$ as given in figure \protect\ref{fig12}.
Since the pulses are left moving pulses, the relation between 
the frequency  $\omega$ and $k$ is given by the 
left-moving branch labeled as such in the figure. Since the final   
pulse is chosen to be a purely negative frequency pulse, it is, as mentioned 
previously, composed of purely positive values of k. Now evolve the wave 
backward in time until one has evolved it past its interaction with the 
horizon and back out into the region where the velocity is a constant. 
Because the problem is time independent, the frequency $\omega$ is conserved. 
Thus the amplitude of the wave over some finite frequency regime must be 
the same.
Furthermore, we have already noted that the left moving wave remains a 
left moving wave, and thus remains on the same left moving branch of 
the dispersion relation. 

This amplitude is given by
\be
\int \tilde\pi^*(t,k) \tilde\phi(t,k) dk
\ee
However, on the left moving branch, we have a definite relation between 
$\omega$ and $k$, and between $\tilde\pi(t,k)$ and $\tilde\phi(t,k)$. In 
particular for the left moving branch, $\tilde\pi(t,k)= +F(k)\tilde\phi(t,k)$.
Thus, we have that the magnitude of the wave over a frequency interval 
$d\omega$ is
\be
\int F(k) |\phi(t,k)|^2 {dk\over d\omega}d\omega
\ee
where the integral is over all values of $k$ for which $\omega(k)$ has 
the desired 
value. From the plot we see that along the left moving branch there are 
three 
values of $k$ for any value of $\omega$ near zero. If we choose our initial 
pulse to have $k$ positive, two of these lie on the negative frequency 
branch, and one lies on the positive frequency branch. Now the initial 
pulse is chosen to be non- zero only for small values of $k$. After the 
interaction with the horizon, the pulse
has been blue shifted, and has non-zero values for $k$ lying near $\pm k_c$,  
the
wave number where the phase velocity of the waves in the constant velocity 
regions ($\omega(k)/k$) is zero. 

Let us define the three values of $k$ at which the value of $\omega$ is 
the same as $k_i$, for the value of k nearest $k=0$, $k_n$ for the nearest 
value of $k$ to $+k_c$ and $k_p=-k_n$ for the value nearest $-k_c$.  Under 
the thermal hypothesis, the initial pulses, obtained by transporting the 
final pulse backward in time, will be composed of positive and negative 
frequency components, with the two magnitudes in a frequency interval related 
by
\be
|F(k_p)\tilde\phi(k_p)^2/v_g(k_p)|
= e^{-\omega(k_i)/T} |F(k_n)\tilde\phi(k_n)^2/v_g(k_n)|
\ee
where $v_g(k)= {d\omega(k)\over dk}$ is the group velocity at wave vector 
$k$.
Furthermore, we should have 
\be
|F(k_p)\tilde\phi(k_p)^2/v_g(k_p)|= {e^{-\omega(k_p)/T} \over 1-e^{-\omega(k_p)/T}} 
|F(k_i) \tilde \phi(k_i)^2/v_g(k_i)|
\ee
Only if these relations hold will the Bogoliubov transformation satisfy 
the thermal hypothesis for all final wave forms. 

\figloc{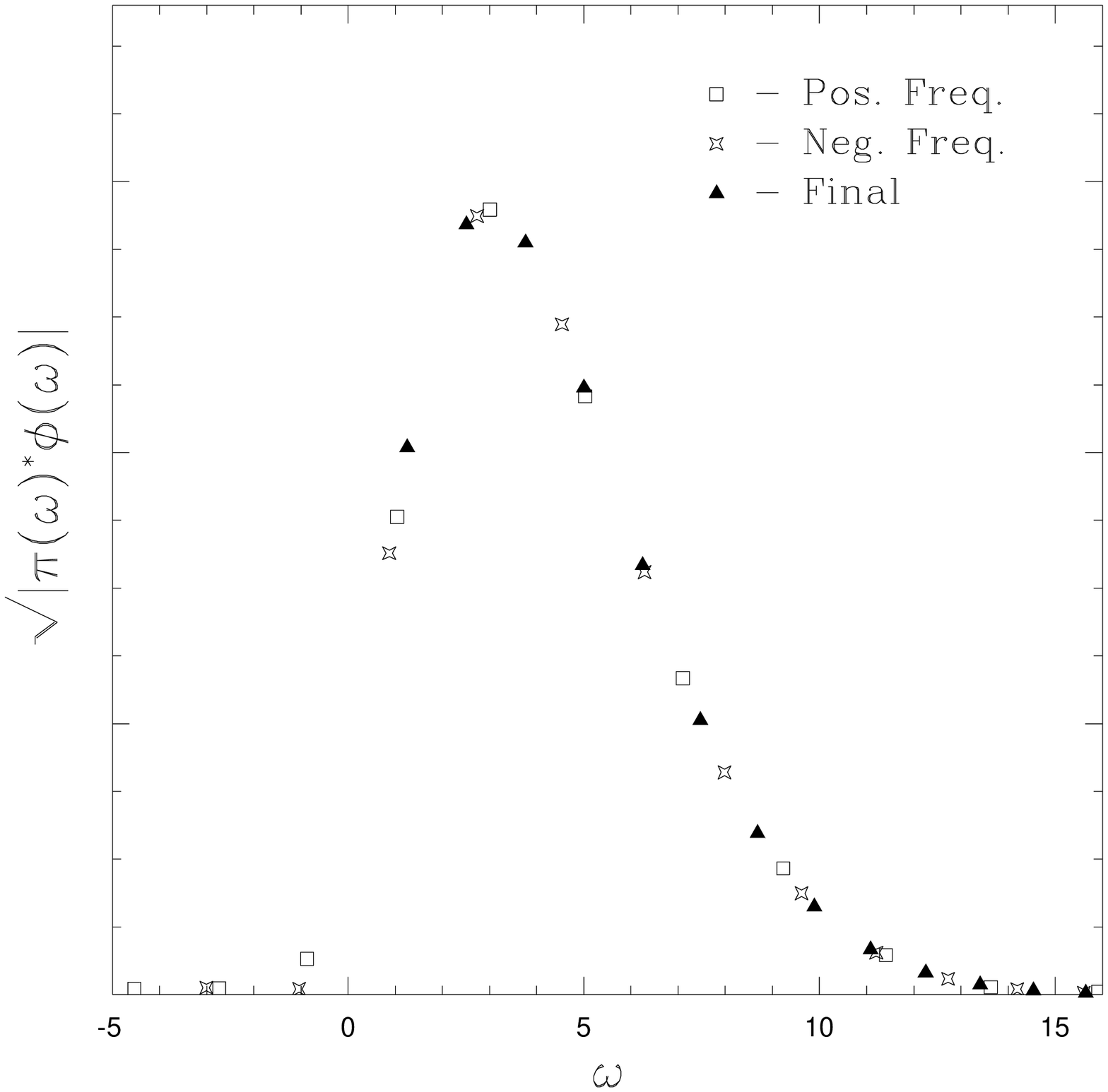}{  The comparison between the thermal predictions for the positive
frequency component of the outgoing pulse and the actual positive frequency 
component of the outgoing pulse.}

In figure \protect\ref{fig13} I have plotted the three functions 
$|F(k_p)\hat\phi_i^+(k_p)^2/v_g(k_p)|^{1/2}$, the positive frequency part of the 
initial pulse, 
$|{e^{-\omega(k_p)/T}
 \over 1-e^{-\omega(k_p)/T}} F(k_i) \hat \phi_f(k_i)^2/v_g(k_i)|^{1/2}$,
the predicted positive frequency part of the inital pulse from the final pulse 
and $|e^{-\omega(k_i)/T} F(k_n)\hat\phi_i^-(k_n)^2/v_g(k_n)|^{1/2}$, the prediction of
the positive frequency part of the final pulse from the negative frequency part 
of the final pulse. These are plotted against the frequency $\omega$.
 These are for the final pulse of figure \protect\ref{fig3} and 
associated initial pulse of figure \protect\ref{fig6}.  Note that 
if the first two agree, the last one's agreeing is a test 
of the conservation of the norm by the evolution.
  Because of the finite size of 
the region and the periodic boundary conditions, only a finite number of 
$k$ values are calculated, but it is clear that these three curves are 
identical to numerical accuracy. Note that the temperature is of order 
1.3, and this
graph thus represents about 10 e-foldings of the thermal function. Thus 
 we can conclude that not only are 
the two particular final wave forms chosen for the soft dispersion relation 
thermal, but that any final wave form would be thermal. Thus despite the 
change in the high frequency dispersion function, the output of the dumb 
hole will continue to be thermal. The  spectrum, and not just the integrated 
intensity, 
 is a thermal spectrum.

 \figloc{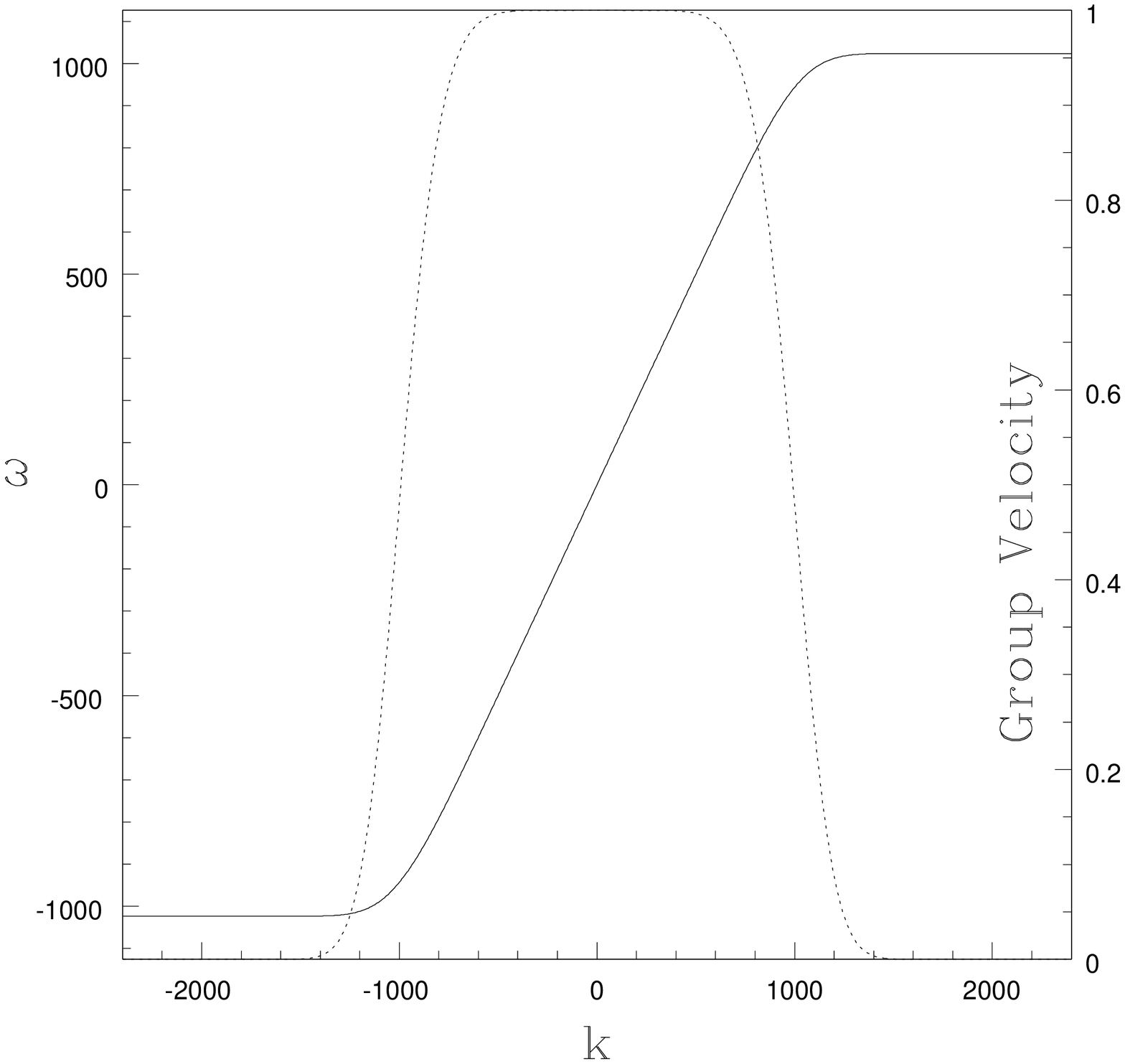}{ The sharp dispersion function ($n=4$) and group velocity (dotted
line) at zero fluid velocity. The transition wave number here is taken 
to be 1024. }

\figloc{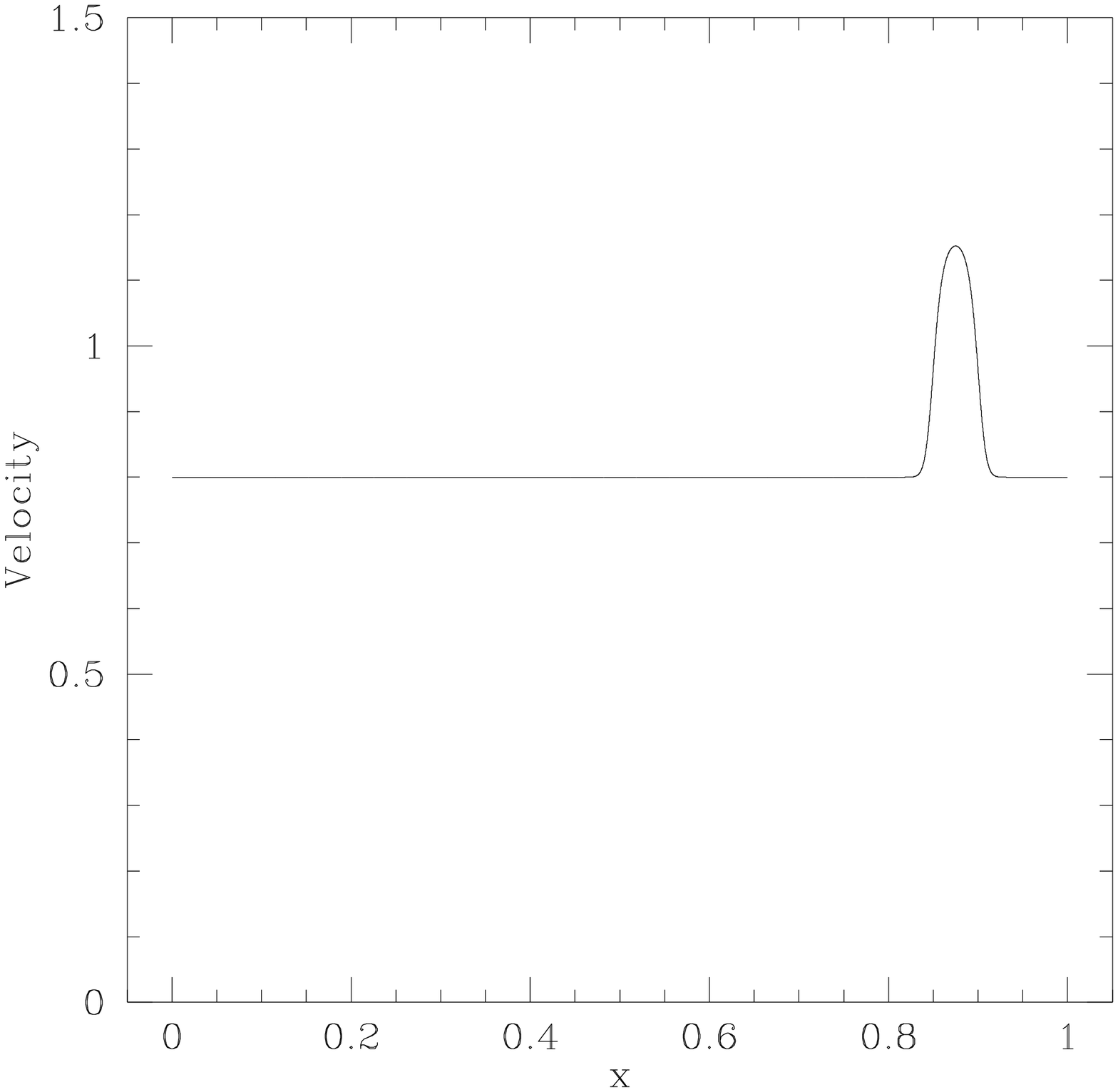}{ The velocity profile for the sharp dispersion function. It 
is scaled by a factor of 2 from that of figure 
\protect\ref{fig1} to allow more room for the 
constant velocity region}

\figloc{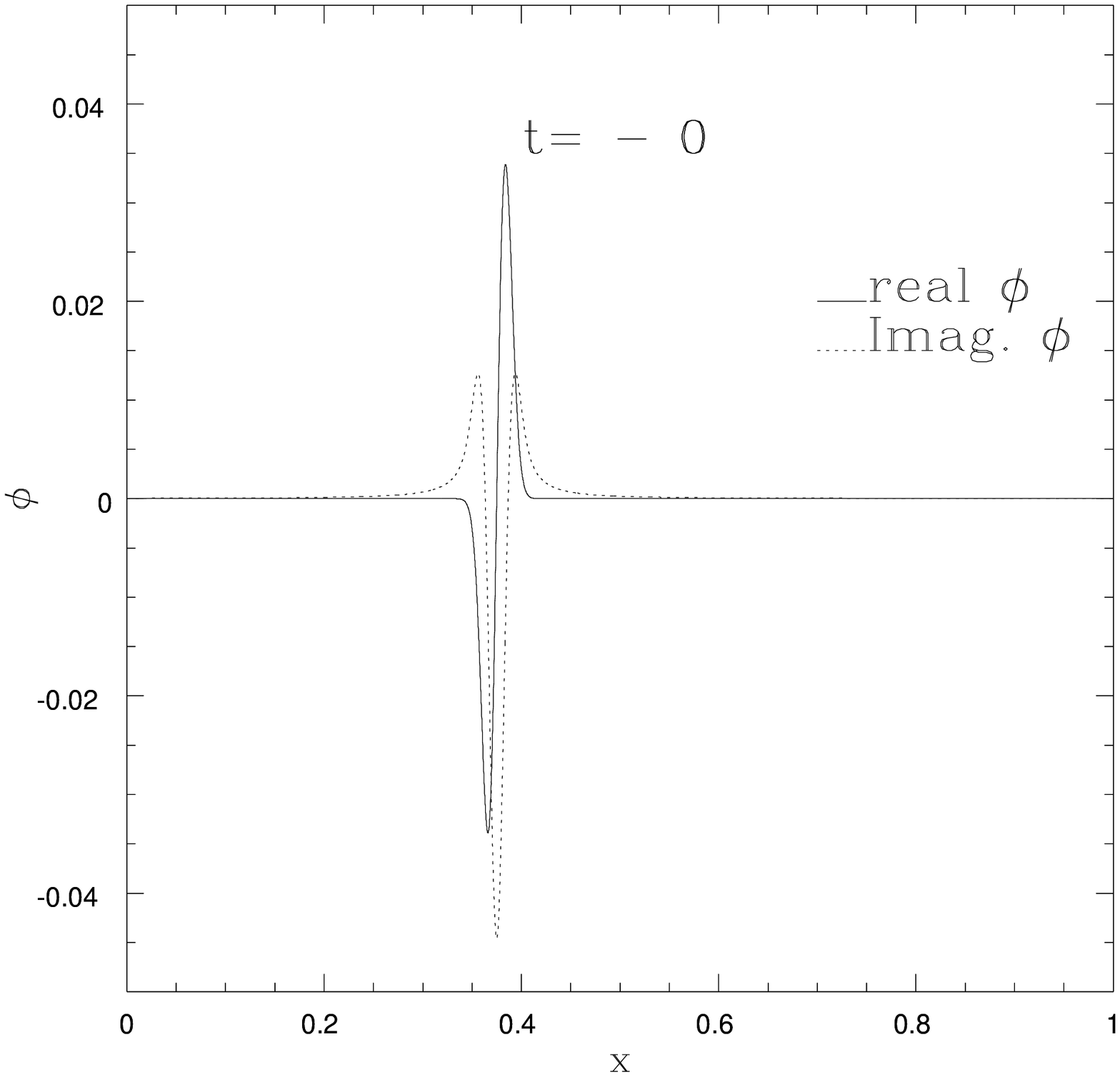}{The final pulse for the sharp dispersion function--- it is scaled
by a factor or 2 from figure \protect\ref{fig3}}

\figloc{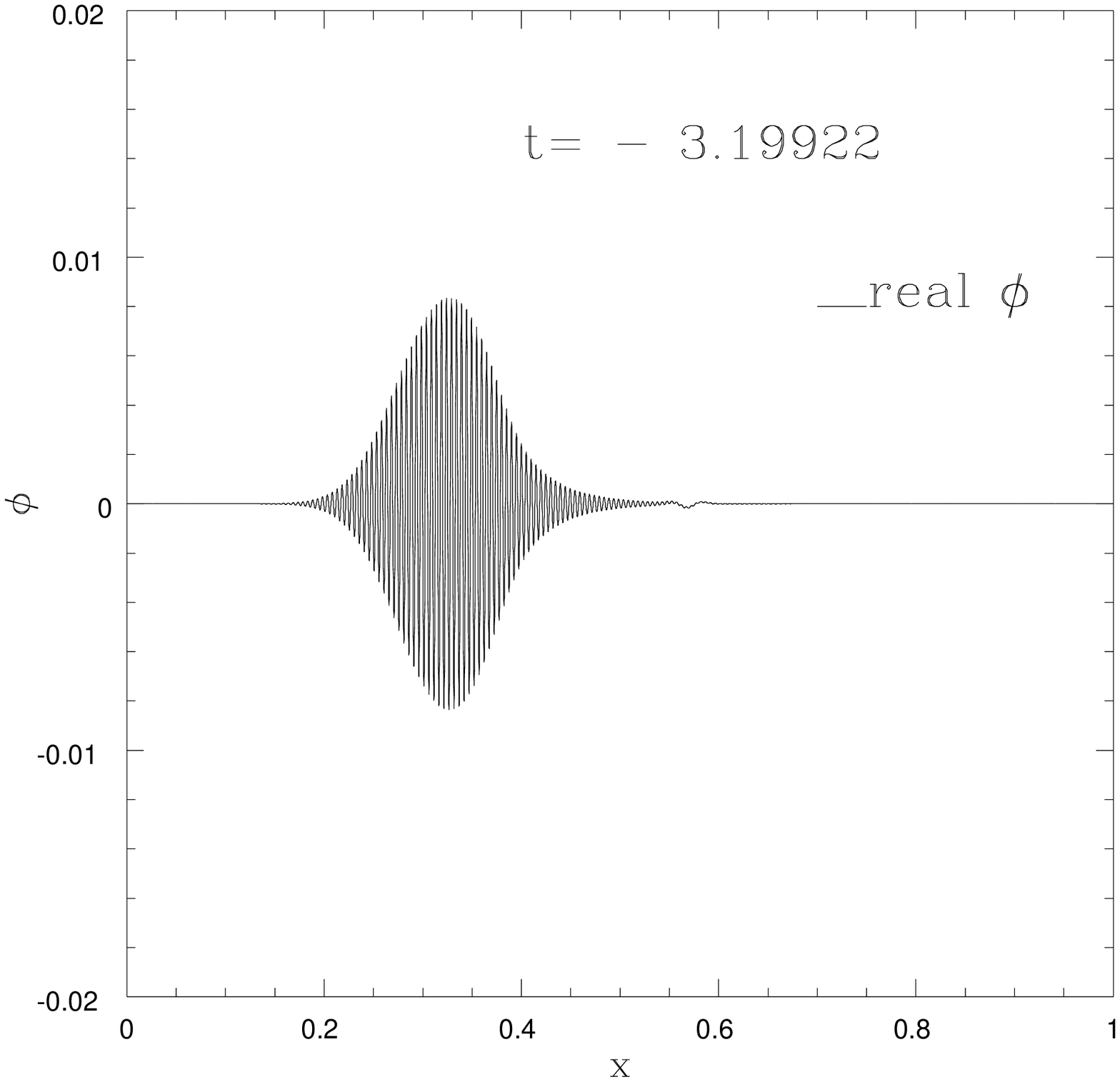}{The inital pulse which would have produced the final pulse 
of figure \protect\ref{fig4}
 rescaled by a factor of 1.5. Note that the initial pulse is more monochromatic 
than that
of figure \protect\ref{fig6} because of the much sharper dispersion function.}

\figloc{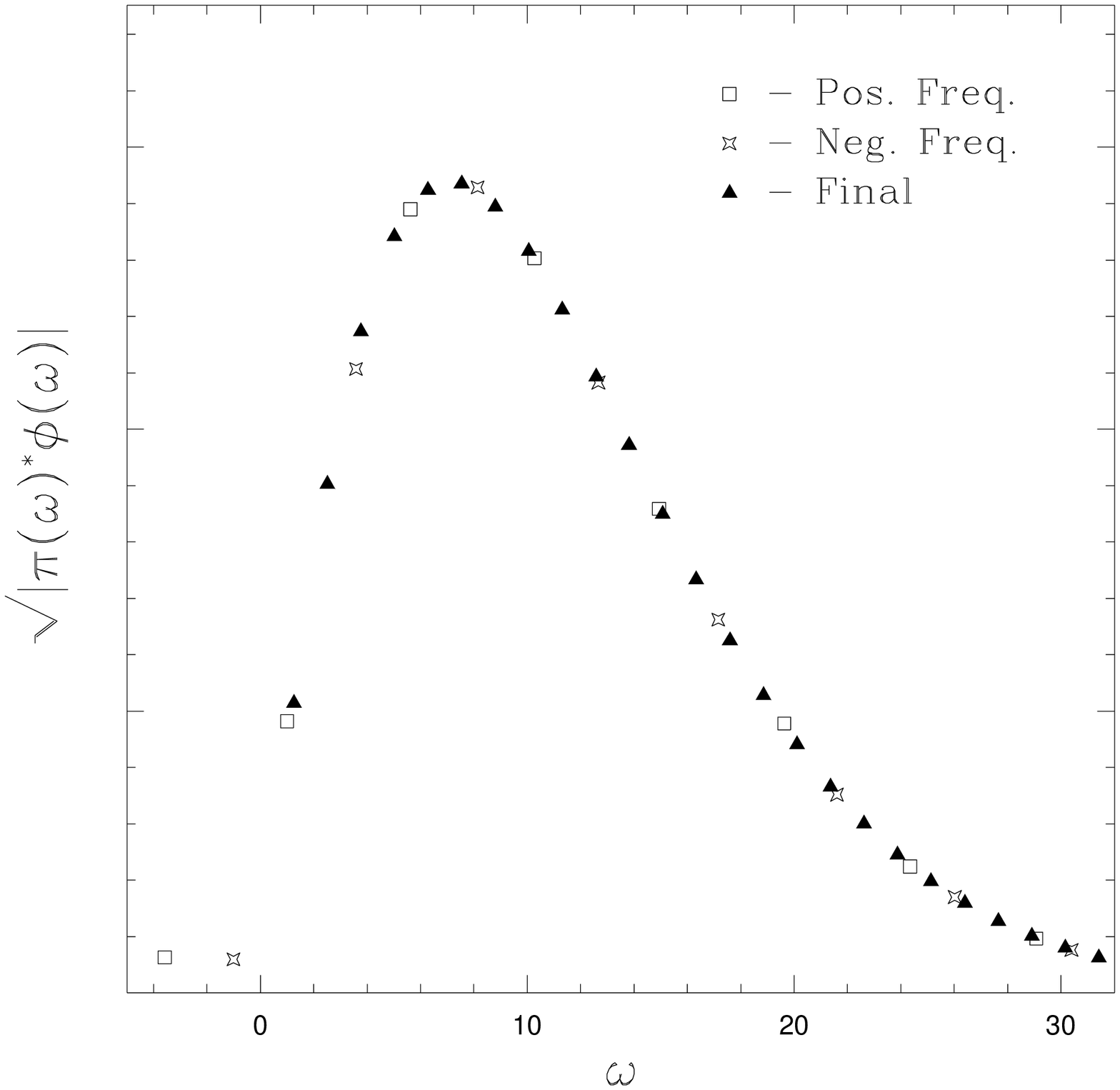}{ The intensity of the positive frequency component of the initial 
pulse of fig. \protect\ref{fig15}
together with the thermally scaled negative frequency spectrum of figure 
\protect\ref{fig16} and that of 
the final pulse of figure \protect\ref{fig3} (rescaled). 
Note that the thermal hypothesis 
does accurately
predict the relations between the various components of the spectrum of 
the ingoing 
and outgoing pulses. }

The dispersion function with $n=1$ present a relatively gentle transition 
from the low
frequency linear regime to the high frequency constant regime. If we alter 
the dispersion 
function to give a much more sudden transition, namely the dispersion function 
with $n=4$ (see fig. \protect\ref{fig14}),
does this alter the thermal spectrum of the emitted radiation? The answer 
is no.
The sharp dispersion function has a much longer initial pulse for the same
situation as used above for the $n=1$ case. In order to allow the initial 
pulse to be contained entirely within the constant velocity region, I have
scaled the problem by a factor of 2. Thus the transition $k_0$ in figure 
\protect\ref{fig14} is larger by 2 to 1024, the velocity profile of figure 
\protect\ref{fig15} has been rescaled by a factor of 2, 
increasing the temperature to approximately 2.7. The 
final pulse of figure \protect\ref{fig16} is just
 figure \protect\ref{fig3} rescaled by a faxtor of 2.
In figure \protect\ref{fig17} I present the initial
 pulse which must have produced the final 
pulse of figure \protect\ref{fig16}
in the case in which $n=4$. Despite the far more sudden transition 
between the regimes in the
dispersion function, the relation between the final and the initial pulses 
is
 again accurately given by a thermal 
relation as plotted in figure \protect\ref{fig18}. 
Again the final positive and negative frequency components of the initial pulse
can be accurately predicted from the final pulse using the thermal hypothesis.

\section{Discussion}

Why are the results thermal, and what lessons can we learn from this simple
problem
for the more interesting one of black hole evaporation? The crucial lessons 
seems
to be that, despite the naive derivations, black hole evaporation depends 
neither
on the ultra high frequency behaviour of the fields near the black hole, 
nor on
the behaviour of the field theory in imaginary time. The process of the 
thermal particle
creation appears to be primarily a low, not high, energy phenomenon.
The black hole and dumb hole have  natural time scales- the the mass in 
the former case, and the inverse gradient of velocity in the latter. The
behaviour of the fields near the horizon of the holes is governed by these 
relatively (compared with Plank or atomic) time scales. In the vicinity 
of
the hole, the fields experience changes on these scales. However,  as far
as the high frequency parts of the field are concerned, these time scales 
are
very long and the changes are slow. The high frequency phenomena simply
adjust adiabatically to these slow changes. If the state is the vacuum 
state at
high frequencies, it remains the vacuum. One would expect this to remain 
true
no matter what the form of the theory was at those high frequencies and 
energies.
It is only in the regime where the natural time scale of the hole is of 
the same
order as the time scale of the field theory that these slow changes near 
the hole
can make their influence felt. I.e., it is only for those
 wave whose frequencies are of
the order of 1/M for the black hole that anything happens.

On that time scale, the structure near the black hole is that the null 
geodesics diverge
exponentially. The time between neigbouring outgoing 
geodesics is exponential in the 
 affine parameter along the geodesics. It is this exponential divergence
 ( shared by both the black hole and the dumb hole) that leads to the
thermal spectrum emitted by the hole. It is this exponential divergence 
at the low frequency
time scale, and not at the highest frequencies of the theory that matters 
for the thermal
emission.

One can encapsulate this freedom of the low frequency thermal
 emission from traces of the high freqeuncy
 behaviour of the theory by the phrase that 
dumb holes (and thus by hypothesis also black holes)  have no
 quantum hair.
Just as the red shift at the horizon wipes out all traces of
 the rich complexity of classical material which could have 
formed the black hole, it also seems to wipe out all traces
 of the quantum structures at high frequencies wich are red 
shifted to form the thermal radiation from the black hole.

\section*{Acknowledgments}
I would like to thank Ted Jacobson who revived my interest in dumb holes, 
and
pointed out in conversations and in ref.\cite{Jacobson} that one of the 
crucial
effects of the atomic nature of matter is to alter 
the high frequency dispersion 
relation of the sound waves. Conversations with him
 have been very helpful to me
in understanding dumb holes.
I would also like to thank Matt Choptuik 
who insisted through the years on
thinking about the stability of the solution scheme
 before coding rather than when the numerical 
instabilities become obvious. Without the lessons
 he taught me   I would not have
solved this problem. In addition a conversation with
 him on the numerical techniques needed for this problem gave me the
courage to attack it. This research was 
carried out under a Fellowship from
 the Canadian Institute for Advanced Research and with an NSERC research 
grant.

\appendix

\section{Numerical Technique}

The equations of motion for the system are solved via a set 
of implicit, symmetric in space and time, finite difference equations. 
The spatial grid 
$\{x_i\}$ is taken as a uniform grid, with $\Delta_x\equiv x_{i+1}-x_i$. The
term $[F(\partial^2_x)\phi](x_i)$ is evaluated by taking the FFT of
 $\phi(x_i)$, $\tilde\phi(k_j)$, multiplying
by $F(-k_j^2)$, and then taking the inverse FFT. The 
temporal grid is staggered, with $\phi_{m,i}=\phi(t_m,x_i)$ defined 
on the time slices $t_i$ with uniform
differences $\Delta_t=t_{m+1}-t_m$. The $\pi$ on the other 
hand are evaluated on the staggered sites
$\pi_{m,i}=\pi(t_m+\half\Delta_t,x_j$. The grid points run from
 0 to $N+1$ where $N=1/\Delta_x$. The periodic boundary conditions are
enforced by taking $ \phi_{m,N}=\phi_{m,0}$ and $\phi_{m,N+1}=\phi_{m,1}$,
and similarly for $\pi$.

The differencing step equations are taken to be
\ba
\phi_{m+1,i} &-& {\Delta_t\over 2\Delta_x}
 \left(v(x_{i+1}\phi_{m+1,i+1} -v(x_{i-1}\phi_{m+1,i-1}\right)
\nen\\
&=&\phi_{m,i}+{\Delta_t\over2\Delta_x} \left( v(x_{i+1}\phi_{m,i+1}
 - v(x_{i-1})\phi_{m,i-1}\right) 
+ \Delta_t \pi_{m,i}
\\
\pi_{m+1,i} &-& {\Delta_t\over 2\Delta_x}v(x_i) \left( \pi_{m+1,i+1} - 
\pi_{m+1,i-1}\right)
\nen\\
&=&
\phi_{m,i}+{\Delta_t\over2\Delta_x}v(x_i) \left(  \pi_{m,i+1} -  \pi_{m,i-1}\right)
 +\Delta_t [F(\partial_x^2) \phi]_{m+1,i}
\ea

This differencing technique was chosen because, if we take $v$ as a constant, 
these equations can be shown to be stable, and non-damping.
 Both of these characteristics
are crucial for the success of the integration. The stability
 is important because of the severe
frequency shifting that takes place on the horizon, a
 frequency shifting which could easily
incite any instabilities present in the code. The non-damping nature is 
also 
crucial because of the necessity of extracting small effects ( the Bogoliubov
coefficients) from the result of the evolution. In particular, since
 we are testing precisely the high frequency effects, it is important not 
to hide
these behind artificial viscosity at high frequencies. 

To see the stability, assume that the velocity is constant, 
and Fourier analyze the difference
equations in both space and time.  We get the characteristic equation
\be
(2i \sin(\omega\Delta_t/2) - iv{\Delta_t\over\Delta_x} \sin(k\Delta_x))^2 
= \Delta_t^2 F(-k^2)
\ee
or
\be
\sin(\omega\Delta_t/2 )= \half\Delta_t\left(\pm\sqrt{-F(-k^2)}
+ {v\over\Delta_x} \sin(2 k \Delta_x)\right)
\ee
This will have real solutions for $\omega$ for all $k$ 
 as long as $\Delta_t\over \Delta_x$ 
is chosen
 to be sufficiently small. In particular the RHS of the equation must be 
less than unity 
for all values of $k$. Since $-F(-k^2)$ is bounded by
 $k_0^2$, and $k_0< k_{max}={\pi\over\Delta_x}$,
and since $\sin(2k\Delta_x)<1$, the right hand side is bounded
 by $\Delta_t \left(k_0 + {v\over \Delta_x}\right)$ so we require 
\be
\Delta_t < {\Delta_x\over k_0\Delta_x+v_{max}}
\ee
in which case the $\omega $ are all real--- the equations are
 stable, and non-damping. This would not have been true had I 
employed an explicit scheme for the time difference equations.

Because of the use of an implicit scheme, 
the time stepping requires the inversion of
matrix equation. To solve this, I used a double sweep 
technique. To illustrate the technique,
let us solve the equation
 \be
 V^0_iU_i - V^p_iU_{i+1} +V^n_iU_{i-1} =S_i \label{defu}
\ee
Both of the equations for the time stepping are of this form, with different 
coefficient vectors
$V^0,V^n,V^p$. These coefficients are the same at all times
as they depend only on $v$ and ${\Delta_t\over\Delta_x}$.
The solution will be found by assuming the recursion relation
\be
U_i=A_i U_{i-1}+B_i
\ee
First solve the recursion relation by substituting for $U_{i+1}$ and $U_i$ 
using this recursion relation into eqn \protect\ref{defu}. Setting the coefficients 
of $U_{i-1}$ to zero, we get recursion relations for  the $A_i$ and $B_i$:
\ba
A_{i}= -{V^n_i\over A_{i+1} V^p_i+V^0_i}
\\
B_{i}= -{V^p_i B_{i+1}\over A_{i+1} V^p_i+V^0_i}
\ea
with $B_{N+1}=1$.
 Now find two solutions to the homogeneous equations
\ba
U^0_{i+1}= {A_{i+1} U^0_i +B_i}
\\
U^1_{i+1}=A_{i+1} U^1_i
\ea
with $U^0_0=0$ and $U^1_0=1$.
Neither of these are periodic.

To find a periodic solution to the non-homogeneous equations, first
recursively solve
\be
B_i= {S_i-V^p_iB_{i+1} \over A_{i+1} V^p_i+V^0_i}
\ee
with $B_{N+1}=0$. Then recursively solve
\be
U_{i}= A_i U_{i-1} +B_i
\ee
with $U_0=0$
This will not satisfy the boundary conditions. However, one can now
subtract a suitable linear combination of $U^0$ and $U^1$ to enforce the
periodic boundary conditions.

 \begin{figure}
\caption{The soft  dispersion function ($n=1$) and corresponding group
velocity (dotted line) at zero fluid velocity. The transition wave number 
$k_0$ is 512.}
\label{fig1}
\end{figure}

\begin{figure} 
\caption{ The velocity profile of the fluid used. The system is
assumed to have periodic boundary conditions. Note that $x\approx .7$
corresponds to the horizon. $x\approx .8$ is a Cauchy horizon.}
\label{fig2}
\end{figure}

\begin{figure}
\caption{ The Real (solid line) and Imaginary (dotted line) parts of a 
final pulse emerging 
from the
dumb hole. The dispersion function has $n=1$ and $k_0=512$.}
\label{fig3}
\end{figure}

 \begin{figure}\caption{ An intermediate stage in the evolution backward in time of the 
real part
of the pulse of figure \protect\ref{fig3}.
 Note that the front of the wave has crashed into 
the horizon,
and been blue shifted to a high frequency pulse whose group velocity is 
less than the fluid
velocity. The dotted line is the same calculation but at half the spatial 
resolution as the solid line.}
\label{fig4}
\end{figure}

\begin{figure}
\caption{ The pulse as it must have been initially to produce the outgoing 
pulse of
figure \protect\ref{fig3}.  Note that it is an
 outgoing pulse but is being dragged in toward 
the horizon by the
inflowing fluid. Again the dotted line is at half the resolution as the 
solid line to give an estimate of the numerical errors.}
\label{fig5}
\end{figure}

\begin{figure}\caption{ The spectrum of the final pulse of
 figure \protect\ref{fig3}. Note that it is 
purely positive wave number
and negative frequency.}
\label{fig6}\end{figure}

\begin{figure}\caption{ The spectrum of the initial 
pulse of figure \protect\ref{fig5}. 
The solid line is positive frequency components
while the dotted line is negative frequencies. 
Since positive frequencies correspond to positive
wave number and vice versa, this still represents a purely outgoing wave.}
\label{fig7}\end{figure}

\begin{figure}\caption{ A narrower final pulse but 
otherwise the same as in Figure \protect\ref{fig3}.}
\label{fig8}\end{figure}

\begin{figure}\caption{ The initial pulse which must have
 produced figure \protect\ref{fig8}.}
 \label{fig9}\end{figure}

\begin{figure}
 \caption{The spectrum of the final pulse of figure \protect\ref{fig8}.}
\label{fig10}\end{figure}

\begin{figure} 
\caption{ The positive and negative frequency spectrum corresponding 
to figure \protect\ref{fig9}. 
 Note the much smaller positive frequency  components than in Figure 
\protect\ref{fig7}. Note that the
power which has not been blue shifted corresponds to components of the 
initial packet whose
group velocity was already of the order of or less than .8, the minimum 
velocity of the
fluid.}
\label{fig11}\end{figure}

\begin{figure}
 \caption{  The relation between $\omega$, the temporal frequency, and
$k$ the wave number in the constant velocity ($V=.8$) region . The various 
branches  (positive and negative frequency and right and left moving) of 
the dispersion relation are labeled.} 
\label{fig12}\end{figure}

\begin{figure}
\caption{  The comparison between the thermal predictions for the positive
frequency component of the outgoing pulse and the actual positive frequency 
component of the outgoing pulse.}
 \label{fig13}\end{figure}

 \begin{figure}\caption{ The sharp dispersion
 function ($n=4$) and group velocity (dotted
line) at zero fluid velocity. The transition wave number here is taken 
to be 1024. }
 \label{fig14}\end{figure}

\begin{figure} \caption{ The velocity profile for
 the sharp dispersion function. It is scaled by a 
factor of 2 from that of figure \protect\ref{fig1} to allow more room for the 
constant velocity region}
\label{fig15}\end{figure}

\begin{figure} \caption{The final pulse for the 
sharp dispersion function--- it is scaled
by a factor or 2 from fig. 3}
\label{fig16}\end{figure}

\begin{figure} \caption{The inital pulse which would have
 produced the final pulse 
of figure \protect\ref{fig16}.}
\label{fig17}\end{figure}

\begin{figure}\caption{ The intensity of the
 positive frequency component of the initial 
pulse of fig. 15
together with the thermally scaled negative 
frequency spectrum of figure 
16 and that of 
the final pulse of figure \protect\ref{fig3} (rescaled). 
Note that the thermal hypothesis 
does accurately
predict the relations between the various components
of the spectrum of 
the ingoing 
and outgoing pulses. }
 \label{fig18}\end{figure}

\end{document}